\documentclass[aps,prl,twocolumn,superscriptaddress,floatfix]{revtex4-2}
\usepackage{graphicx}
\usepackage{amsmath}
\usepackage{amssymb}
\usepackage{bm}
\usepackage{braket}
\usepackage{dcolumn}
\usepackage{orcidlink}
\usepackage{hyperref}
\usepackage[normalem]{ulem}

\begin{document}

%Title of paper
\title{Anomalous temperature dependence of polaron mobility in a nonlinear double-well potential: unbiased X-propagator approach}

\author{Stefano~Ragni\,\orcidlink{0009-0003-5603-2968}}
\email{sragni@ifs.hr}
\affiliation{Department for Research of Materials under Extreme Conditions, Institute of Physics, 10000 Zagreb, Croatia}
\author{Osor~S.~Bari\v{s}i\'c\,\orcidlink{0000-0002-6514-9004}}
\affiliation{Department for Research of Materials under Extreme Conditions, Institute of Physics, 10000 Zagreb, Croatia}
\author{Naoto~Nagaosa}
\affiliation{RIKEN Center for Emergent Matter Science (CEMS), Wako, Saitama 351-0198, Japan}
\affiliation{Fundamental Quantum Science Program (FQSP), TRIP Headquarters, RIKEN, Wako 351-0198, Japan}
\author{Andrey~S.~Mishchenko\,\orcidlink{0000-0002-7626-7567}}
\email{andrey.mishchenko6363@gmail.com}
\affiliation{Department for Research of Materials under Extreme Conditions, Institute of Physics, 10000 Zagreb, Croatia}
\affiliation{RIKEN Center for Emergent Matter Science (CEMS), Wako, Saitama 351-0198, Japan}

\begin{abstract}
We develop an unbiased X-propagator method for calculating finite-temperature optical conductivity $\sigma(\omega)$ and dc mobility $\mu$ for arbitrary nonlinear electron–phonon interaction. We apply it to a polaron coupled to a double-well lattice potential, a minimal model for strongly anharmonic polar materials. At moderate coupling, the mobility exhibits three temperature regimes associated with confinement within one well, thermal competition with the barrier, and barrier-insensitive high-temperature dynamics. This sequence produces a concave temperature dependence of the mobility that is absent in conventional linear-coupling polaron models. At strong coupling, the mobility becomes nonmonotonic, and its temperature evolution is reflected in a characteristic redistribution of optical spectral weight. For parameters relevant to SrTiO$_3$, our results qualitatively reproduce both the anomalous concave mobility and the onset of violation of the Mott–Ioffe–Regel limit, thereby supporting nonlinear coupling to a soft anharmonic lattice mode as a microscopic mechanism for anomalous transport in dilute polar metals.
\end{abstract}

\maketitle

%%% INTRODUCTION

Charge transport in dilute polar metals and oxides has long exhibited behavior that, over a wide temperature range, deviates from that expected from transport theories based on standard scattering mechanisms \cite{BehniaReview2019, Sch2021}. One of the prototypical systems is doped SrTiO\textsubscript{3}, where the mobility $\mu$ shows an unusual temperature dependence that persists over several decades in temperature and carrier density from $10^{17}$ to $10^{19}$ cm$^{-3}$ ~\cite{Tufte, BehniaNPJ2017, BehniaReview2019, BehniaScience, BehniaMIRPRX}. A compound with similar behavior of the resistivity  $R \sim \mu^{-1}$ is KTaO\textsubscript{3} \cite{Wemple, Tokura}. The low-temperature mobility follows $T^{-2}$ behavior, which cannot be explained by Umklapp processes of the Fermi liquid theory due to the small Fermi surface in the dilute limit \cite{BehniaNPJ2017, BehniaReview2019}, but has received a plausible explanation in terms of quadratic scattering from transverse optical phonons \cite{Maslov2021}. Furthermore, at $T\approx 50$ K, the temperature dependence gradually crosses over to a puzzling $T^{-n}$ law with $n\approx3$, which has been discussed in the context of first-principles calculations \cite{BernardiPRL} involving soft phonon modes and anharmonic lattice dynamics. These transport regimes have been studied extensively experimentally, and it is well established that at elevated temperatures with $T^{-3}$ law the Mott--Ioffe--Regel (MIR) limit is violated, with the carrier mean free path becoming shorter than the interatomic spacing \cite{BehniaNPJ2017, BehniaReview2019, BehniaMIRPRX}. While individual aspects of experimentally observed behavior have been addressed by different theoretical approaches, there is no satisfactory theoretical framework that consistently accounts for the experimental observations across the entire $T^{-2}$--$T^{-3}$ range, where the temperature dependence of the mobility is concave.

The success of explaining the low-temperature $T^{-2}$ and $T^{-3}$ dependencies using different, yet in both cases phonon-related, approaches \cite{Maslov2021, Bernardi} raises the possibility that the unusual transport behavior originates more generally from electron--phonon interactions (EPI) \cite{Gunnar, Maslov2021, Mishchenko2015, Mishchenko2019, Sch2021, Bernardi}. These interactions dress charge carriers with lattice distortions, forming quasiparticles known as polarons. 
Depending on the coupling strength and phonon spectrum, polarons may exhibit either coherent band-like transport or incoherent hopping, giving rise to a wide variety of temperature dependences of the mobility \cite{Mishchenko2015, Mishchenko2019}. 
However, standard polaron models with linear EPIs fail to reproduce the transport behavior observed in dilute oxides \cite{Sch2021, BehniaReview2019}. In particular, approximation-free calculations of the mobility \cite{Mishchenko2015, Mishchenko2019} show that, except in the unlikely strong-coupling regime, the rapid decrease of the mobility at low temperatures crosses over to a slower decrease at higher temperatures, leading to the well-known resistivity saturation \cite{Gunnar}. 
By contrast, the mobility of SrTiO\textsubscript{3} exhibits the opposite trend: around $50$ K, the temperature dependence evolves from a weaker to a stronger power law, approximately $T^{-2} \rightarrow T^{-3}$ \cite{BehniaNPJ2017, BehniaReview2019, Maslov2021}. To our knowledge, no existing model reproduces the experimentally observed concave temperature dependence of the mobility, even at a qualitative level. Moreover, the relation between this $T^{-3}$ mobility regime and the subsequent violation of the MIR limit remains unresolved.

Recent developments suggest that polaron formation in many complex materials is more accurately described by nonlinear EPIs \cite{Sch2021}. Examples include doped manganites \cite{manganites,Hoesch2013}, halide perovskites \cite{Sch2021, Saidi2016, Zacharias2023}, and quantum paraelectrics \cite{Bilz, Maslov2021, Nazaryan2021, Ranalli2024}. In particular, a natural minimal model of lattice anharmonicity is provided by a double-well potential \cite{Spaldin}, which is relevant to both SrTiO\textsubscript{3} and KTaO\textsubscript{3}. Such a potential can arise from the combination of a positive quartic ($g_4>0$) and a negative quadratic ($g_2<0$) electron--phonon coupling and was first investigated theoretically in the context of intercalated sublattices of heavy and light atoms \cite{Adolphs2014, AdolphsEPL, Adolphs2}. For $g_2<0$, the quartic term stabilizes the lattice and allows the exploration of a broad range of negative quadratic couplings \cite{stefano23}. The resulting interaction introduces several characteristic energies and, consequently, scattering channels operating on different time scales, which may substantially renormalize polaron transport. Such a rich scattering landscape gives rise to temperature-dependent mobilities that differ qualitatively from those predicted by harmonic \cite{Mishchenko2015, Mishchenko2019} or purely quadratic \cite{stefano23} EPI models.

Our aim is to clarify how strong anharmonicity modifies transport regimes and whether it can account for the anomalous mobility observed in systems such as doped SrTiO\textsubscript{3} and KTaO\textsubscript{3}. To this end, we develop a new unbiased X-propagator approach to calculate the finite-temperature current--current correlation function for arbitrary linear and nonlinear EPIs by extending the zero-temperature formalism of Ref.~\cite{Ragni2025}.
Analytic continuation then yields the approximation-free frequency-dependent mobility $\mu(\omega)$, equivalently the optical conductivity per particle per charge 
$
\mu(\omega)=\sigma(\omega)/(ne)
$, 
as well as the static dc mobility $\mu\equiv\mu(\omega=0)$. Because the method is formulated for arbitrary lattice potentials, it is readily applicable to a broad class of anharmonic systems.

%%% MODEL
{\it Model and parameters.}
Intrinsic anharmonicity is known to play an important role in considered materials, introducing both bare and interaction-induced nonlinearities \cite{Spaldin}. To reduce the number of parameters, we consider the limiting case of a harmonic bare lattice and retain only nonlinear terms generated through the EPI~\cite{Adolphs2014, Ragni2025}. This choice enables us to isolate the transport consequences of carrier-induced lattice nonlinearities without introducing unnecessary details of the underlying lattice dynamics. The polaron Hamiltonian in the lattice displacement ($x$) representation is
\begin{equation}
\widehat{H} = -t \sum_{\langle ij \rangle} c_i^{\dagger} c_j + 
\sum_j -\frac{1}{2} \frac{\partial^2}{\partial x_j^2} + V(x_j) \;,
\label{h-gen}    
\end{equation}
with potential
\begin{equation}
V(x_j) = \frac{\Omega^2}{2} x_j^2 + c_j^{\dagger} c_j^{\,} \sum_{n=2,4} g_n (2\Omega)^{n/2} x_j^n \;.
\label{loc3}    
\end{equation}
Here $c_j$ ($c_j^{\dagger}$) are annihilation (creation) operators on site $j$, $x_j$ is the local lattice displacement operator, $t$ is the electron hopping, $\Omega$ is the phonon frequency, and $g_n$ are EPI constants for the corresponding nonlinearity $n$. We study a case identical to one of those considered in \cite{Ragni2025}, i.e., the one-dimensional model with $t=1$, $\Omega=0.25$, $g_4=0.1$, and various negative values of $g_2$, allowing the consideration of double-well potentials with different regimes of barrier heights. 
The Boltzmann constant k$_{\rm B}$, lattice spacing $a$, Planck constant $\hbar$, and charge $e$ are set to unity. 

In addition to $\Omega$, the potential (\ref{loc3}), being double-well if $g_2<-\Omega/4$, is characterized by the barrier height 
$
W = (4g_2/\Omega + 1)^2 \Omega^2/64g_4 \; ,
$
and vibrational frequency in one of the double wells $\tilde{\omega}=\sqrt{2\Omega(-4g_2-\Omega)}$
 \cite{Adolphs2014, Ragni2025}. To probe all regimes we consider $g_2=-0.45$ ($W \approx 3\Omega/2$), $g_2=-0.96$ ($W \approx 3\tilde{\omega}/2$), and additional values $g_2 = -0.2, -0.8, -1.4, -1.8$ chosen to interpolate between the reference points. For the considered $\Omega=0.25$, all $g_2$ values correspond to a double-well potential.

%%% RESULTS

\begin{figure}
    \centering
    \includegraphics[width=\linewidth]{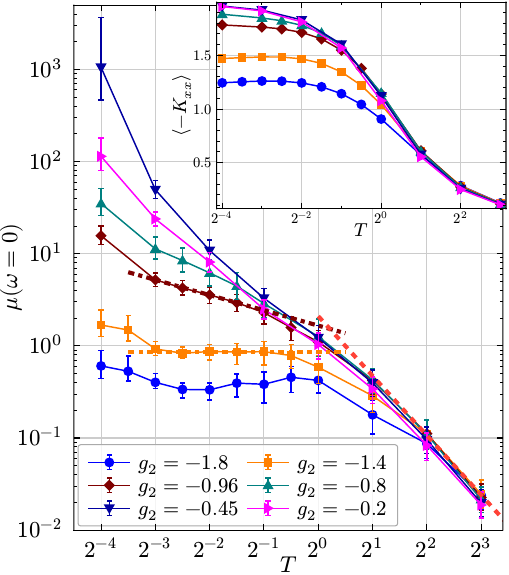}
    \caption{Temperature dependence of the static mobility $\mu$ and kinetic energy $\langle - K_{xx} \rangle$ (inset) for $g_4=0.1$, $\Omega=0.25$, and different quadratic coupling constants. The dash-dotted brown (orange) line highlights the middle-temperature trend for $g_2 = -0.96$ ($g_2=-1.4$), while the dashed bright red line highlights the high-temperature trend shared by all couplings. The error bars in the inset are smaller than the dot size.}
    \label{fig:mobility}
\end{figure}

{\it Results}. Figure \ref{fig:mobility} shows the mobilities $\mu$ and kinetic energies $\langle -K_{xx} \rangle$ for different values of $g_2$, corresponding to different regimes of the barrier $W$. In comparison with the harmonic potential case \cite{Mishchenko2015, Mishchenko2019}, one first observes that the temperature dependence of the mobility is fundamentally different: it is concave instead of convex in the moderate and strong coupling regimes. Furthermore, Fig.~\ref{fig:mobility} shows that the larger the temperature, the smaller the dependence of the mobility and the kinetic energy on the height of the barrier.  The $g_2=-0.2$ curve corresponds to an almost zero barrier $W<0.05  \ll \Omega$ and can serve as a reference for the non-barrier case. The larger the absolute value of $g_2$, corresponding to a larger barrier height, the higher the temperature at which the corresponding quantity converges to the reference non-barrier value. 

Qualitatively, the temperature dependence of the mobility falls into three regimes. 
The values $g_2 \ge -0.45$ belong to the weak coupling non-barrier regime with the quasi-particle residue $Z \gtrapprox 0.93$ \cite{Ragni2025} and barrier height $W \le 3\Omega/2$. In this case, the temperature dependence is rather featureless, with a gradual decrease as the temperature rises and a faster drop in mobility at lower temperatures than at higher ones. This is similar to the typical behavior of barrierless single-well linear EPI models for weak coupling \cite{Mishchenko2015, Mishchenko2019}.

\begin{figure}
    \centering
    \includegraphics[width=0.9\linewidth]{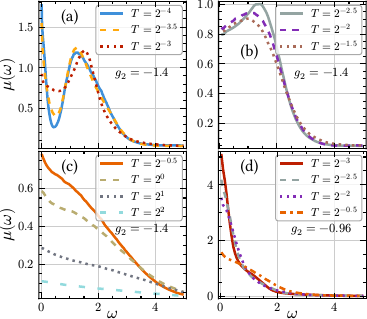}
    \caption{Optical conductivity spectra at various temperatures for $g_2 = -1.4$ (a,b,c) and $g_2 = -0.96$ (d). $g_4 = 0.1$ in all plots.}
    \label{fig:spectra}    
\end{figure}

Parameters $g_2 = -0.8$ and $g_2=-0.96$ belong to the moderate coupling, i.e., the intermediate height barrier regime with $0.6 \lessapprox Z  \lessapprox 0.85$ \cite{Ragni2025} and $3\Omega/2 < W \le 3\tilde{\omega}/2$. 
This is precisely the regime in which the temperature dependence of the mobility is concave.
Moreover, the double-well case features three temperature regimes, in contrast to the two regimes of the linear case \cite{Mishchenko2015, Mishchenko2019}. For temperatures much smaller than the barrier, a fast drop at low temperatures $T \le 2^{-3}$ is associated with a single well of the double-well potential.
The middle temperature range $2^{-3} \lessapprox T \lessapprox 2^{-1}$ is associated with the interplay between the temperature and the barrier, because, at the high temperature end of this domain, the mobility approaches the reference non-barrier curve. Finally, for high temperatures $T \gtrapprox 2^{-1}$, the system is insensitive to the presence of the barrier. 
The curve for $g_2=-0.96$ demonstrates these three domains most clearly, being highlighted in Fig.~\ref{fig:mobility} for the middle- and high-temperature domains by dash-dotted and dashed lines, respectively.   

For $g_2 = -1.4$, the system lies at the boundary between moderate and strong coupling regimes with $Z \approx 0.15$ \cite{Ragni2025} and $W \approx 2 \times  3\tilde{\omega}/2$, and it has the same three temperature domains as the $g_2 = -0.8$ and $g_2 = -0.96$ cases. 
In Fig.~\ref{fig:mobility}, the middle-temperature range is highlighted by a dash-dotted line. 
Of particular interest is the close correspondence between the temperature domains and the shape of the frequency-dependent mobility $\mu(\omega)$, as shown in Fig.~\ref{fig:spectra}. The low temperature domain (Fig.~\ref{fig:spectra}a) corresponds to a double peak $\mu(\omega)$ with peaks at $\omega=0$ and $\omega>0$. The middle temperature domain is characterized by a single peak at $\omega \ne 0$ (Fig.~\ref{fig:spectra}b), and $\mu(\omega)$ has a single broad peak at $\omega=0$ for high temperatures (Fig.~\ref{fig:spectra}c). It is clear that such correspondence is a feature of the strong coupling regime because the case of $g_2=-0.96$ has no such correspondence (Fig.~\ref{fig:spectra}d), whereas the strong coupling case with $g_2=-1.8$ does (see the Supplemental Material \cite{Supp}).         

The $g_2=-1.8$ case belongs to the strong-coupling (large-barrier) regime, with $Z\approx0.03$ \cite{Ragni2025} and $W\gg3\tilde{\omega}/2$. As in the case of linear EPI \cite{Mishchenko2015,Mishchenko2019}, the mobility in the strong-coupling regime exhibits a nonmonotonic temperature dependence. In particular, this behavior is observed in the Holstein \cite{Mishchenko2015}, Fröhlich \cite{Mishchenko2019,Frost}, and double-well models, suggesting that it is a general feature of a broad class of polaronic systems.

{\it Anomalous transport in SrTiO\textsubscript{3}}. At moderate coupling, the double-well model captures the key qualitative features of the temperature-dependent resistivity observed in SrTiO\textsubscript{3}~\cite{BehniaNPJ2017,BehniaReview2019,Maslov2021}. 
A quantitative comparison is beyond the present modeling since SrTiO\textsubscript{3} is a three-dimensional system rather than one-dimensional. Although the ground state of the nonlinear quadratic interaction is insensitive to the dimension of the system \cite{stefano23}, the temperature dependence can change due to a different density of states. Besides, the actual soft-phonon branch is strongly dispersive~\cite{Covley,Stirling}, whereas the model (\ref{h-gen}, \ref{loc3}) assumes dispersionless phonons. 
In addition, the frequency of the Brillouin zone center $\Gamma_{15}$ transverse optical phonon increases with temperature \cite{Covley, Stirling}, modifying the Hamiltonian (\ref{h-gen}, \ref{loc3}). 
However, none of the above differences can alter the general trend of concave behavior in the double-well model, because the main properties of the temperature dependence of the mobility arise from the interplay between the energy of excited phonons and the barrier height.
Indeed, any peculiarities in the density of states and phonon dispersion can be outweighed by the larger barrier height.  
Qualitatively, for parameters close to their experimental values, the model reproduces both the characteristic concave temperature dependence of the mobility and the violation of the MIR limit at high temperatures, where the mobility decreases most rapidly. 

The experimentally relevant regime lies between $g_2=-0.96$ and $g_2=-1.4$, for which the two corresponding power-law behaviors are highlighted in Fig.~\ref{fig:mobility}. Our theoretical estimate of the mass enhancement yields $m^*/m\approx1.6$ for $g_2=-0.96$ and $m^*/m\approx5.8$ for $g_2=-1.4$ \cite{Ragni2025}. These values bracket the experimental estimate $m^*/m\lesssim3$ reported for carrier concentrations between $10^{17}$ and $10^{19}$~cm$^{-3}$ \cite{mass1,massMcCalla}. 

\begin{figure}
    \centering
    \includegraphics[width=0.85\linewidth]{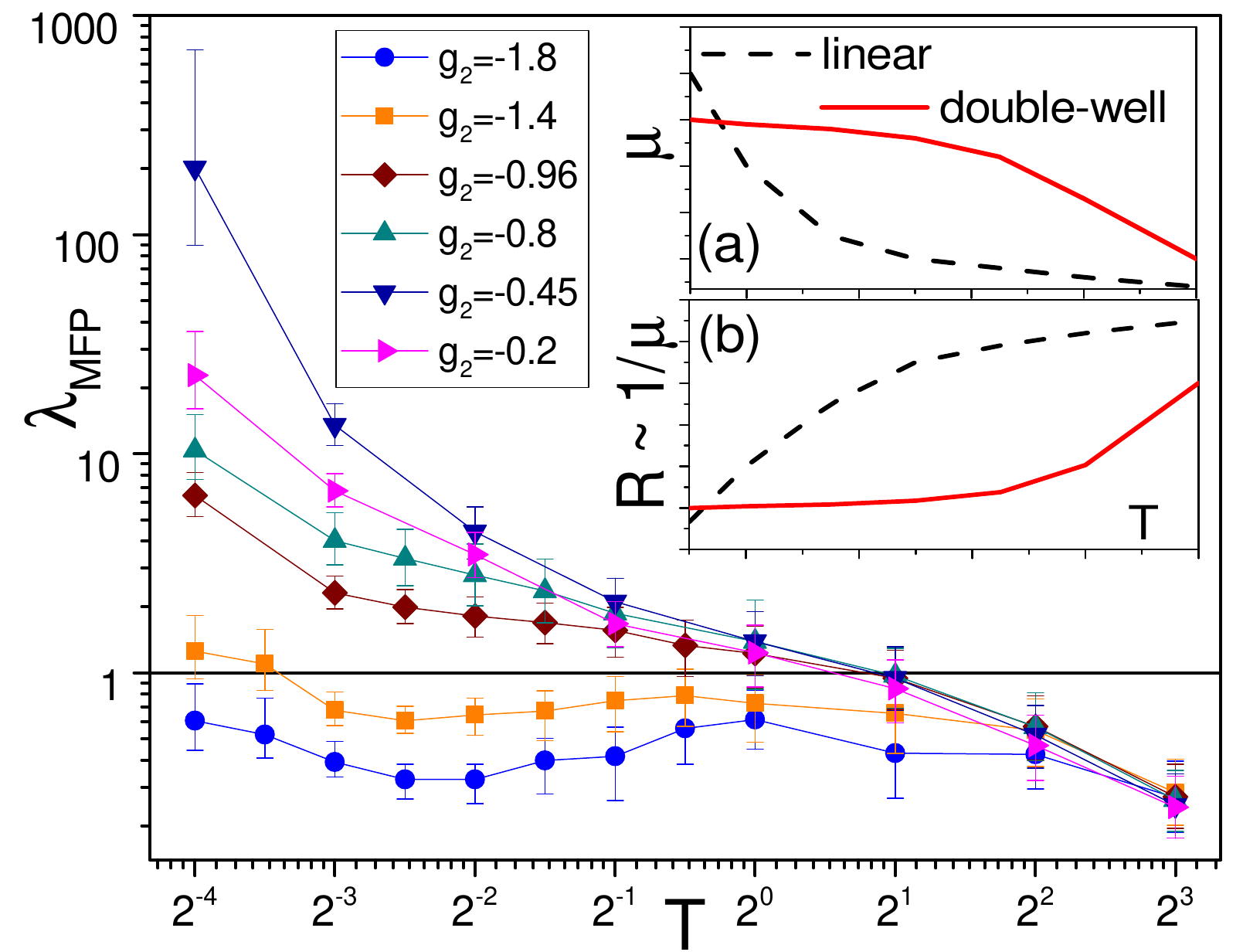}
    \caption{Temperature dependence of the mean free path $\lambda_{\rm MFP}$ in units of lattice spacing $a$ for different quadratic coupling constants $g_2$. Inset: schematic temperature dependence of the mobility and resistivity for the linear (dashed line) and double-well (solid line) models.}
    \label{fig:MFP}
\end{figure}

The mobility in linear EPI models typically exhibits a convex temperature dependence, as illustrated in inset (a) of Fig.~\ref{fig:MFP}: a rapid decrease upon lowering the temperature at low $T$ is followed by a weaker temperature dependence at higher $T$. Consequently, the resistivity $R\propto1/\mu$ rises rapidly at low temperatures and subsequently approaches the well-known saturation regime \cite{Gunnar}, see inset (b) of Fig.~\ref{fig:MFP}. In contrast, over a broad temperature range, the double-well model exhibits a slower decrease in mobility in the middle temperature domain and a faster drop as the temperature rises, being consistent with the experimental concave dependence \cite{BehniaNPJ2017, BehniaReview2019}. Indeed, the resulting resistivity behavior is consistent with that observed in SrTiO\textsubscript{3} \cite{Maslov2021}. 

A further key result is that the model predicts a violation of the MIR limit, in direct agreement with experiment. The mean free path $\lambda_{\rm MFP}$ can be estimated as
$
\lambda_{\rm MFP}\approx \mu\sqrt{C_{JJ}(\tau=0)}/\langle K_{xx}\rangle,
$
where $C_{JJ}(\tau)$ is the current–current correlation function in imaginary time \cite{mori1,mori2,Golez,Mishchenko2015}. As shown in Fig.~\ref{fig:MFP}, for the experimentally relevant parameter range, $g_2\lesssim-0.96$, the estimated mean free path falls below the lattice spacing $a=1$, i.e., violates the MIR limit, upon entering the temperature regime in which the mobility begins to decrease more rapidly. This almost simultaneous onset of accelerated mobility suppression and MIR-limit violation qualitatively reproduces the behavior observed experimentally in SrTiO\textsubscript{3} \cite{BehniaNPJ2017,BehniaReview2019,BehniaMIRPRX}. Taken together, the agreement with a broad range of experimental observations strongly indicates that carrier coupling to a highly nonlinear soft lattice provides a microscopic mechanism for the anomalous transport behavior of SrTiO\textsubscript{3}.

%%% METHOD

{\it Method}. Our current technique extends the Numerical X-Propagators method introduced in \cite{Ragni2025} for the imaginary-time electron Green’s function to the generation of partition function diagrams. The main differences are: (i) periodic boundary conditions are imposed, the electron must be on the same site at $\tau=0$ and $\tau=\beta$. (ii) The configuration now has a fixed length on the $\tau$ axis, given by $\beta = 1/T$.

\begin{figure}
    \centering
    \includegraphics[width=0.9\linewidth]{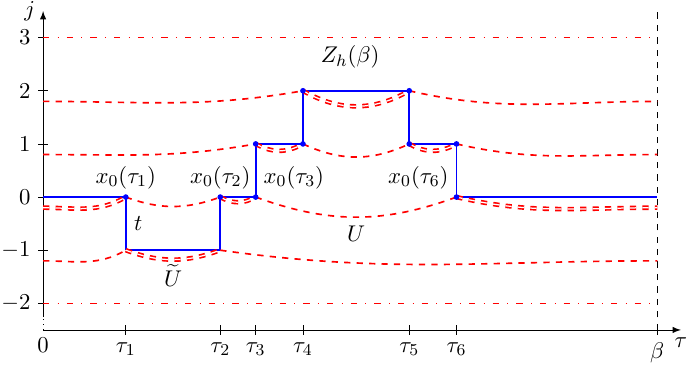}
    \caption{A representative configuration generated by the algorithm. The horizontal axis denotes the imaginary times $\tau$ of the hopping events, while the vertical axis labels the lattice sites $j$. Events are represented as blue points, and they are associated with an oscillator displacement $x_j(\tau_a)$. The electron trajectory is shown as a solid blue line, where vertical lines represent hoppings $t$ between sites $\Delta j$. The time propagation of oscillators is depicted in red: a single dashed line corresponds to an unoccupied site propagator $U$, a double dashed line to an occupied site propagator $\widetilde U$.}
    \label{fig:partition-function-technique}
\end{figure}

Figure~\ref{fig:partition-function-technique} illustrates a representative configuration generated by the algorithm. 
The function $U(y,x,\tau)$ describes the propagation of the oscillator in the absence of the electron ($c_j^{\dagger} c_j^{\,} = 0$ in Eq.~\ref{loc3}), and the Monte Carlo weight of sites not reached by the electron is
$
Z_h(\beta) = \int {\rm d}x \, U(x,x,\beta) = (1-e^{-\beta\Omega})^{-1}.
$
The propagator $\widetilde U(y,x,\tau)$ is used in the presence of the electron ($c_j^{\dagger} c_j^{\,} = 1$ in Eq.~\ref{loc3}) \cite{Ragni2025}. The generation of configurations starts from the atomic limit diagram with zero hoppings and follows a Metropolis scheme with the following elementary updates:

\textit{Add kink-antikink.} One electron propagator is selected at random. A pair of opposite hoppings $\Delta j$ and $-\Delta j$ (kink-antikink) is introduced along the selected propagator, so that the electron path on the rest of the diagram is preserved.

\textit{Remove kink-antikink.} One hopping is selected at random. If the next hopping is its inverse, i.e. together they form a kink-antikink pair, the pair may be removed while satisfying detailed balance.

\textit{Update $x$.} One event is selected at random, and its oscillator displacement $x$ is changed.

\textit{Update $\tau$.} One hopping is selected at random, and the imaginary time $\tau_a$ at which it occurs is changed without alteration of the diagram topology.

A more detailed description of the updates is provided in the Supplemental Material~\cite{Supp}.

Our technique provides direct access to the imaginary-time current--current correlation function
$
C_{JJ}(\tau) = \left\langle
\mathcal{T}_{\tau}\hat{J}(\tau)\hat{J}(0)
\right\rangle ,
$
where the current operator associated with the kinetic term in Eq.~\eqref{h-gen} is
$
    \hat J = -it \sum_j (c^\dag_{j+1}c^{\,}_j - c^\dag_{j-1}c^{\,}_j).
$
Each sampled partition-function diagram contains a variable number $N$ of hopping events at times $\tau_a$, with directions $\Delta j_a= \pm 1$. Replacing a hopping vertex by a current insertion multiplies its diagrammatic weight by $-i\Delta j_a$. Every ordered pair of distinct hopping events can be used as a pair of current insertions, yielding
\begin{equation}
C_{JJ}(\tau)
=
-\frac{1}{\beta}
\left\langle
\sum_{a\neq b}^{N}
\Delta j_a\,\Delta j_b\,
\delta_\beta\!\left(\tau-\tau_a+\tau_b\right)
\right\rangle_{\mathrm{MC}}.
\label{eq:imaginary-time-corr}
\end{equation}
where $\delta_\beta(x) = \sum_{m\in\mathbb{Z}}\delta(x-m\beta)$
is the $\beta$-periodic delta function, and
$\langle\cdots\rangle_{\mathrm{MC}}$ denotes the normalized Monte Carlo
average over partition-function diagrams. The factor $1/\beta$ normalizes
the average over the arbitrary position of the imaginary-time origin.

To eliminate imaginary-time binning errors, we measure the Matsubara
correlator directly. Using the convention
\begin{equation}
C_{JJ}(i\omega_n)
=
\int_0^\beta d\tau\,
e^{i\omega_n\tau}C_{JJ}(\tau),
\qquad
\omega_n=\frac{2\pi n}{\beta},
\end{equation}
Eq.~\eqref{eq:imaginary-time-corr} gives
\begin{equation}
C_{JJ}(i\omega_n)
=
-\frac{1}{\beta}
\left\langle
\sum_{a\neq b}^{N}
\Delta j_a\,\Delta j_b\,
\cos\!\left[\omega_n(\tau_a-\tau_b)\right]
\right\rangle_{\mathrm{MC}}.
\label{eq:matsubara-corr}
\end{equation}

Finally, the mobility spectrum $\mu(\omega)$ is reconstructed from the Matsubara correlators~(\ref{eq:matsubara-corr}) via the Stochastic Optimization Method~\cite{Mishchenko2000, Goulko2017, Krivenko} by inverting the relation
\begin{equation}
C_{JJ}(i\omega_n) = \frac{2}{\pi}
\int_0^{\infty} \!{\rm d}\omega\; \frac{\omega^2}{\omega^2+\omega_n^2} \mu(\omega) \; .
\end{equation}
The static mobility is obtained as $\mu(\omega \to 0)$.

The method was benchmarked against Diagrammatic Monte Carlo for the linear Holstein~\cite{Mishchenko2015} and purely quadratic~\cite{stefano23} models, reproducing the same Matsubara correlators within statistical error.

{\it Conclusions}. We extended the unbiased $X$-propagator approach developed in Ref.~\cite{Ragni2025} to calculations of the optical conductivity and finite-temperature mobility for an arbitrary form of electron--phonon interaction. We applied the method to a double-well potential, which is prototypical of polar metals and oxides, and directly compared the resulting transport behavior with experimental observations in SrTiO\textsubscript{3}.

At low temperatures, the mobility depends strongly on the height of the double-well barrier, whereas at sufficiently high temperatures the different cases converge because the system becomes insensitive to the detailed shape of the potential. In the strong-coupling regime, the mobility exhibits a nonmonotonic temperature dependence, similar to that observed in linear electron--phonon models. This suggests that nonmonotonic mobility may be a general feature of strongly coupled polaronic systems. 

For the moderate-coupling regime most relevant to experiments, the temperature dependence of the mobility comprises three distinct domains. At low temperatures, the carrier remains predominantly confined to one of the two wells, and the mobility decreases rapidly with increasing temperature. At intermediate temperatures, the decrease becomes slower because the thermal energy is comparable to the barrier height. At still higher temperatures, the mobility again drops rapidly as the carrier dynamics become increasingly insensitive to the barrier. For strong couplings, we also find a clear correspondence between the temperature dependence of the mobility and the frequency dependence of the optical conductivity. At low temperatures, the latter displays a two-peak structure, with one peak at $\omega=0$ and another at finite frequency. In the intermediate-temperature domain, only the finite-frequency peak remains, whereas at high temperatures the spectral weight involves a broad structure centered at $\omega=0$.

Our results qualitatively reproduce the anomalous temperature dependence observed in SrTiO\textsubscript{3}, in particular the concave, rather than convex, mobility profile. Moreover, the model predicts a violation of the Mott--Ioffe--Regel limit in the regime of rapid high-temperature mobility suppression, in direct agreement with experiment. These findings link the anomalous transport of SrTiO\textsubscript{3} to the interplay between thermal fluctuations and the double-well structure of a strongly anharmonic soft mode, highlighting nonlinear lattice dynamics as an essential ingredient in dilute polar metals.

{\it Acknowledgments}. We thank N. Prokof'ev for fruitful discussions. This work was supported by the Croatian Science Foundation under the project numbers IP-2024-05-2406 and IP-2022-10-9423, as well as by the Project FrustKor, financed by the EU through the National Recovery and Resilience Plan 2021-2026. N.N. was supported by JSPS KAKENHI Grant Numbers 24H00197, 24H02231 and 24K00583 and by the RIKEN TRIP initiative.

\bibliography{bibliography}

\end{document}

\title{Supplemental material for ``Anomalous temperature dependence of polaron mobility in a nonlinear double-well potential: unbiased X-propagator approach''}

\author{Stefano~Ragni\,\orcidlink{0009-0003-5603-2968}}
\email{sragni@ifs.hr}
\affiliation{Department for Research of Materials under Extreme Conditions, Institute of Physics, 10000 Zagreb, Croatia}
\author{Osor~S.~Bari\v{s}i\'c\,\orcidlink{0000-0002-6514-9004}}
\affiliation{Department for Research of Materials under Extreme Conditions, Institute of Physics, 10000 Zagreb, Croatia}
\author{Naoto~Nagaosa}
\affiliation{RIKEN Center for Emergent Matter Science (CEMS),
Wako, Saitama 351-0198, Japan}
\affiliation{Fundamental Quantum Science Program (FQSP), TRIP Headquarters, 
RIKEN, Wako 351-0198, Japan}
\author{Andrey~S.~Mishchenko\,\orcidlink{0000-0002-7626-7567}}
\email{andrey.mishchenko6363@gmail.com}
\affiliation{Department for Research of Materials under Extreme Conditions, Institute of Physics, 10000 Zagreb, Croatia}
\affiliation{RIKEN Center for Emergent Matter Science (CEMS),
Wako, Saitama 351-0198, Japan}

\maketitle

\section{Optical conductivity spectra in the deep double-well regime}

\begin{figure}[h]
    \centering
    \includegraphics[width=\linewidth]{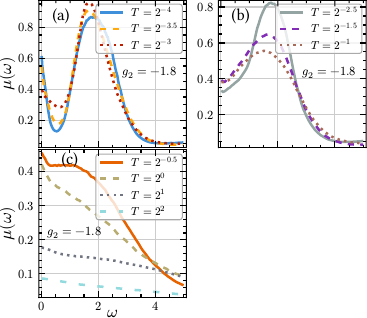}
    \caption{Optical conductivity spectra at various temperatures for $g_2 = -1.8$ and $g_4 = 0.1$.}
    \label{fig:spectra_1.8}
\end{figure}

Fig.~\ref{fig:spectra_1.8} shows the spectra for $g_2 = -1.8$ in the three temperature regimes (a-c), analogous to those in Fig.~2(a-c) of the main text. Again, the low temperature domain (Fig.~\ref{fig:spectra_1.8}a) corresponds to a double peak $\mu(\omega)$ with peaks at $\omega=0$ and $\omega>0$, the middle temperature domain is characterized by a single peak at $\omega \ne 0$ (Fig.~\ref{fig:spectra_1.8}b), and $\mu(\omega)$ has a broad peak at $\omega=0$ for high temperatures (Fig.~\ref{fig:spectra_1.8}c). At the stronger coupling $g_2 = -1.8$, static mobility is reduced compared to $g_2 = -1.4$, whereas the incoherent peak becomes more pronounced. At this coupling, we observe the onset of nonmonotonic behavior of the dc mobility, with the mobility at $T = 2^{-2.5}$ being slightly smaller than at both the lower temperature $T = 2^{-3}$ and the higher temperature $T = 2^{-1.5}$.

\newpage

\section{Description of the updates}

\subsection{Add/remove kink--antikink}

\begin{figure}[ht]
  \centering
  \begin{minipage}{0.9\linewidth}
    \centering
    \includegraphics[width=\textwidth]{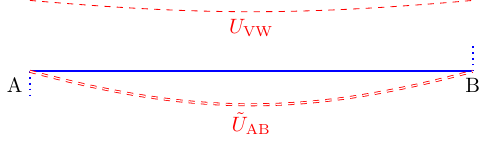}
    $\uparrow\downarrow$\vspace{0.1cm}
    \includegraphics[width=\textwidth]{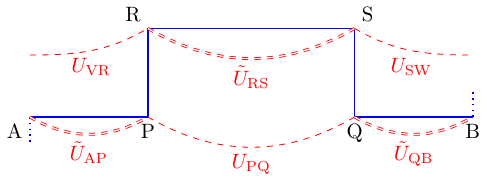}
  \end{minipage}
  \caption{Graphical representation of the kink--antikink update in case (A). Here, a pair of hopping transitions is inserted.}
  \label{fig:update_add}
\end{figure}

One electron propagator is randomly selected from the entire diagram, and its ends are labeled A and B (Fig.~\ref{fig:update_add}). A new hopping transition PR (the ``kink'') is proposed at a time $\tau_{\rm P} = \tau_{\rm R}$, sampled uniformly between $\tau_{\rm A}$ and $\tau_{\rm B}$. To keep the electron's final site fixed, we also propose the insertion of the ``antikink'' SQ, directed opposite to PR, at a time $\tau_{\rm Q} = \tau_{\rm S}$, sampled uniformly between $\tau_{\rm P}$ and $\tau_{\rm B}$. In the complementary inverse update, a random hopping transition is selected, and we check whether it forms a kink--antikink pair with the next transition before proposing their removal.

Let us denote as site $i$ the site of events A and B, and as site $f$ the site of $R$ and $S$. The total Metropolis acceptance ratio is the product of the diagram weight ratio on site $i$, the weight ratio on site $f$, and the probability distribution from which the new coordinates are extracted,
\begin{equation}
    R_{\rm add} = \frac{t^2 \, R_i \, R_f}{\mathcal{P}_{\rm add}} .
\end{equation}

The weight ratio on site $i$ is always
\begin{equation}
    R_i = \frac{\tilde U_{\rm AP} \, U_{\rm PQ} \, \tilde U_{\rm QB}}{\tilde U_{\rm AB}} ,
\end{equation}
provided that we introduce an artificial trace-cut on the electron path at $\tau = 0 \equiv \beta$ so that events $A$ and $B$ can always be identified, even when no hoppings are present.

The weight ratio at site $f$ depends on whether site $f$ has been visited previously (A) or not (B).
\begin{equation}
    R_f = \tilde U_{\rm RS} \times \begin{cases}
    U_{\rm VR} \, U_{\rm SW} / U_{\rm VW} & {\rm (A)}\\
    U_{\rm SR}^{\rm ext} / Z_{\rm h}(\beta) & {\rm (B)}
    \end{cases} ,
\end{equation}
where V and W are events already present on site $f$ in the diagram. In case (B), $Z_{\rm h}(\beta) = (1-e^{-\beta\Omega})^{-1}$ and $U_{\rm SR}^{\rm ext}$ denotes the propagator that wraps around the $\beta$-cylinder, whereas $\tilde U_{\rm RS}$ is the direct propagator between S and R. Lastly, the proposal-probability factor is given by
\begin{align}
    \mathcal{P}_{\rm add} &= \frac{1}{2} \frac{N_{\rm hop}+2}{N_{\rm el}} 
    \frac{1}{\tau_{\rm B} - \tau_{\rm A}} \frac{1}{\tau_{\rm B} - \tau_{\rm P}} \\
    &\times \mathcal{N}(x_{\rm P},x_{\rm Q};\bm{\mu}_{\rm PQ},\bm{\Sigma}_{\rm PQ}) \, \mathcal{N}(x_{\rm R},x_{\rm S};\bm{\mu}_{\rm RS},\bm{\Sigma}_{\rm RS}) .
\end{align}
The factor $1/2$ originates from the two possible choices for nearest neighbor hopping in 1D. The factor $(N_{\rm hop}+2) / N_{\rm el}$ arises from selecting one of the $N_{\rm el}$ electron propagators in the diagram with uniform probability, and selecting one kink--antikink pair for removal in the inverse update. The new $x$-coordinates of the events P and Q added on the same site are sampled from a bivariate Gaussian distribution; its parameters are chosen according to the following procedure, which depends on the local environment around the inserted kinks. Let us define the auxiliary functions
\begin{align}
    a(\tau) \coloneqq (2 \tanh&(\Omega \tau))^{-1} \\
    b(\tau) \coloneqq -(2 \sinh&(\Omega \tau))^{-1} .
\end{align}
An optimized choice for the precision matrix $\bm{\Lambda} = \bm{\Sigma}^{-1}$ for the purely harmonic case is
\begin{gather}
    \lambda_{11} = a(\Delta\tau_l) + a(\Delta\tau_m) , \\
    \lambda_{22} = a(\Delta\tau_r) + a(\Delta\tau_m) , \\
    \lambda_{12} = \lambda_{21} = b(\Delta\tau_m) + \begin{cases}
        0 & {\rm (A)} \\
        b(\Delta\tau_l) & {\rm (B)}
    \end{cases} ,
\end{gather}
$\bm{\mu} = \bm{0}$ in Case (B), while
\begin{equation}
    \bm{\mu} = \bm{\Sigma} \begin{bmatrix}
        -x_l b(\Delta\tau_l) \\
        -x_r b(\Delta\tau_r)
    \end{bmatrix}
\end{equation}
in Case (A). This choice of Gaussian parameters was found to maintain good acceptance rates for the values of $g_2$ and $g_4$ considered in this work.

Specific realizations of the environment are: for site $f$, Case (A), $\Delta\tau_l = \tau_{\rm R}-\tau_{\rm V}$, $\Delta\tau_m = \tau_{\rm S}-\tau_{\rm R}$, $\Delta\tau_r = \tau_{\rm W}-\tau_{\rm S}$, $x_l = x_{\rm V}$, and $x_r = x_{\rm W}$. In Case (B), $\Delta\tau_l = \Delta\tau_r = \beta - (\tau_{\rm S}-\tau_{\rm R})$, $\Delta\tau_m = \tau_{\rm S}-\tau_{\rm R}$. Site $i$ has the same structure as site $f$, Case (A).

\subsection{Change $x$ coordinate}

\begin{figure}[ht]
\centering
\includegraphics[width=0.7\linewidth]{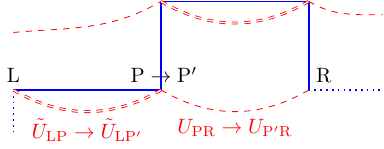}
\caption{Graphical representation of the update which changes the $x_{\rm P} \to x_{\rm P'}$ coordinate of the P event.}
\label{fig:x-chg-x}
\end{figure}

This update changes the $x$ coordinate associated with an event P chosen randomly with uniform probability from all the events in the diagram.

Let us label L the event directly to the left of P, and R the
one directly to its right.  When the coordinate of event P
is changed, the values of the two propagators on the segments LP and PR also
change. It is clear that one of them must always be a $\tilde U$ propagator; in
Fig.~\ref{fig:x-chg-x} we choose the one on the LP segment. The other one is a $U$ propagator.

The new $x_{\rm P'}$ coordinate for the event P is sampled from the normal distribution
$\mathcal{N}(x_{\rm P'},\mu,\sigma)$ with parameters
\begin{align}
    1/\sigma^2 &= a(\Delta\tau_l) + a(\Delta\tau_r) \label{eq:gauss-sigma} , \\
    \mu/\sigma^2 &= -(x_{\rm L} b(\Delta\tau_l) + x_{\rm R} b(\Delta\tau_r)) \label{eq:gauss-mu} ,
\end{align}
where $\Delta\tau_l = \tau_{\rm P} - \tau_{\rm L}$ and $\Delta\tau_r = \tau_{\rm R} - \tau_{\rm P}$. The Metropolis ratio for this update is
\begin{equation}
	R_{\rm chx} = \frac{\tilde U_{\rm LP'} \, U_{\rm P'R}}{\tilde U_{\rm LP} \, U_{\rm PR}}
	\times \frac{\mathcal{N}(x_{\rm P},\mu,\sigma)}{\mathcal{N}(x_{\rm P'},\mu,\sigma)} .
\end{equation}

\subsection{Change $\tau$ of hopping transition}

\begin{figure}[ht]
  \centering
  \begin{minipage}{0.8\linewidth}
    \centering
    \includegraphics[width=\textwidth]{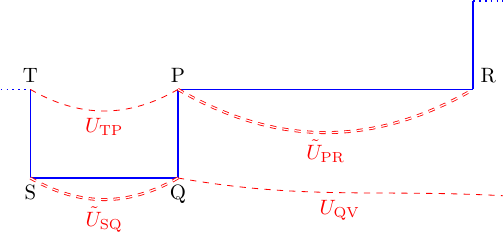}
    $\uparrow\downarrow$\vspace{0.1cm}
    \includegraphics[width=\textwidth]{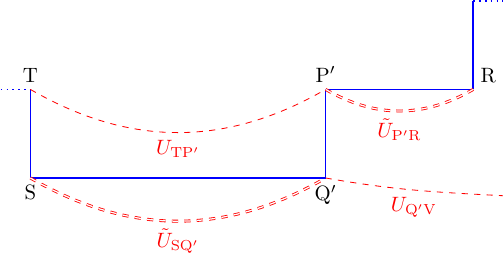}
  \end{minipage}
  \caption{Graphical representation of the update which changes the time of
  one of the hopping transitions in the diagram. The time of events P and Q $\tau_{\rm P}=\tau_{\rm Q}$ is changed to $\tau_{\rm P'}=\tau_{\rm Q'}$.}
  \label{fig:x-chg_tau_i}
\end{figure}

This update randomly selects one hopping transition (which we label PQ), and attempts to modify the imaginary
times $\tau_{\rm P}=\tau_{\rm Q} \to \tau_{\rm P'}=\tau_{\rm Q'}$ at which it occurs. The proposed
$\tau_{\rm P'}$ is extracted from the uniform probability distribution $U(\tau_{\rm P'})
= 1 / (\tau_{\rm R} - \tau_{\rm T})$, see Fig.~\ref{fig:x-chg_tau_i} for the labeling of the vertices. In such a way, the new time cannot go beyond
the bounds imposed by the neighboring interactions, and their order is never
changed.

The Metropolis acceptance ratio for this update is 
\begin{equation}
	R_{\rm ch\tau} = \frac{\tilde U_{\rm P'R}}{\tilde U_{\rm PR}} \frac{\tilde U_{SQ'}}{\tilde U_{\rm SQ}} \frac{U_{\rm TP'}}{U_{\rm TP}} \frac{U_{\rm Q'V}}{U_{\rm QV}} .
\end{equation}